\documentclass{svmult}

\usepackage{makeidx}         % allows index generation
\usepackage{graphicx}        % standard LaTeX graphics tool
\usepackage{multicol}        % used for the two-column index
\usepackage[bottom]{footmisc}% places footnotes at page bottom
\usepackage{bm}
\makeindex             % used for the subject index
\begin{document}

\title*{Influence of particle shape on shear stress in granular media}

\author{Emilien Az\'ema \inst{1}, Farhang Radja\"\i \inst{1}, Robert Peyroux\inst{1}\and
Gilles Saussine\inst{2}}
\institute{LMGC, CNRS - Universit\'e Montpellier II, Place Eug\`ene Bataillon, 
34095 Montpellier cedex 05, France.
\texttt{azema@lmgc.univ-montp2.fr}
\and Innovation and Research Departement of SNCF, 45 rue de Londres, 
75379 PARIS Cedex 08 \texttt{gilles.saussine@sncf.fr}}
\maketitle

We analyze the contact and  force networks in a dense confined packing 
of pentagonal particles simulated by means of the contact dynamics method.  
The particle shape effect is evidenced by comparing the 
data from pentagon packing and from a packing with identical characteristics except for 
the circular shape of the particles. 
A surprising observation  is that the pentagon packing develops a
lower structural anisotropy than the disk packing.
We show in this work that this weakness  is  
compensated by a higher force anisotropy that leads to enhanced shear strength of 
the pentagon packing. With the polygonal shape of the particles,   
the strong force chains are mostly composed of edge-to-edge contacts with a  
marked zig-zag aspect.

\section{Introduction}
\label{sec:1}

The microstructure of granular materials is generically anisotropic in two respects: 
1) The contact normal directions are not random; 2) The force average as 
a function of  contact normal direction is not uniform. 
The corresponding fabric and force anisotropies in shear are responsible 
for mechanical strength at the scale of the packing \cite{Kruyt1996,Bathurst1988,Rothenburg1989,Radjai1998}.  
The shear stress is fully transmitted via a "strong'' contact network, materialized by 
force "chains" and the stability is ensured by the antagonist role 
of  "weak'' contacts which prop strong force chains\cite{Radjai1998,Staron2005}. 
These features have, however,  been for the most part investigated  
 in the case of granular media composed of 
isometric (circular or spheric) particles. 

In this paper, we consider one of the simplest possible shapes, namely 
regular pentagons. Among regular polygons, the 
pentagon has the lowest number of 
sides, corresponding to the least roundedness in this category, without the 
pathological space-filling properties of triangles and squares. 
We seek to isolate the effect of edge-to-edge contacts on shear stress and force transmission 
by comparing the data with a packing of circular particles that, apart from the particle shape, is 
fully identical  (preparation, friction coefficients, particle size distribution) to the pentagon packing. 
Both packings are subjected to biaxial compression simulated by means 
of the contact dynamics method.  The presence  of edge-to-edge contacts 
affects both quantitatively and qualitatively the microstructure and 
the overall behavior during shear. These contacts do not transmit torques, 
but they are able to accommodate force lines that are usually 
unsustainable in packings of disks.    

\section{Numerical procedures}
\label{sec:2}

The simulations were carried out by means of the contact dynamics (CD) method with 
pentagonal particles. 
The CD method is based on implicit time integration of the equations 
of motion and a nonsmooth formulation of  
mutual exclusion and dry friction between particles \cite{Jean1999,Jean1992,Moreau1994}.  
This method requires no elastic repulsive potential and no smoothing 
of the Coulomb friction law 
for the determination of forces. For this reason, the simulations can be 
performed with large time steps compared to molecular dynamics simulations.  
We used LMGC90 which is a multipurpose software  
developed in our laboratory, capable of modeling a collection of deformable or undeformable particles of
various shapes by different algorithms \cite{DUBOIS2003}.

\begin{figure}
\centering
\includegraphics[width=5.5cm]{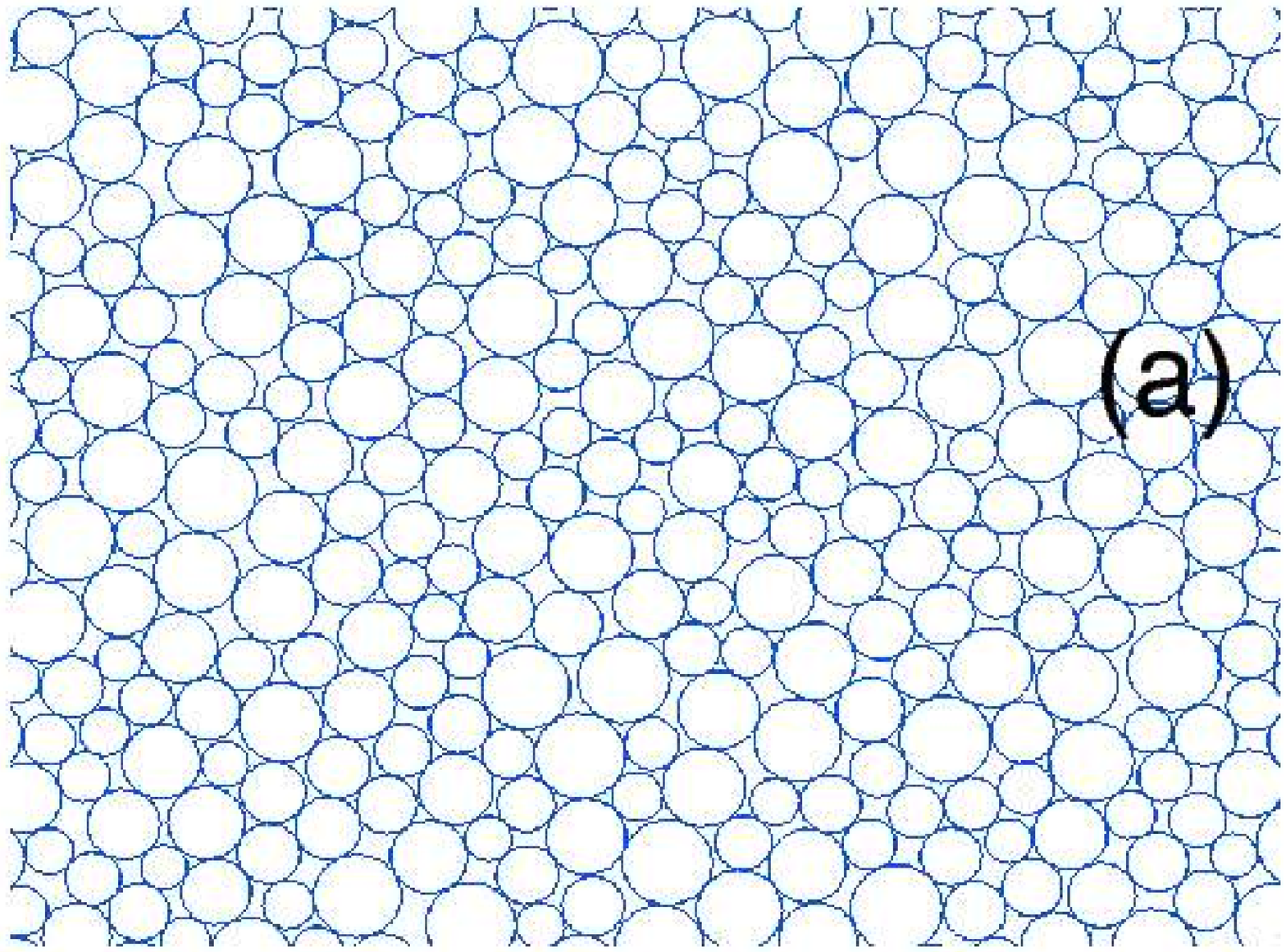}
\includegraphics[width=5.5cm]{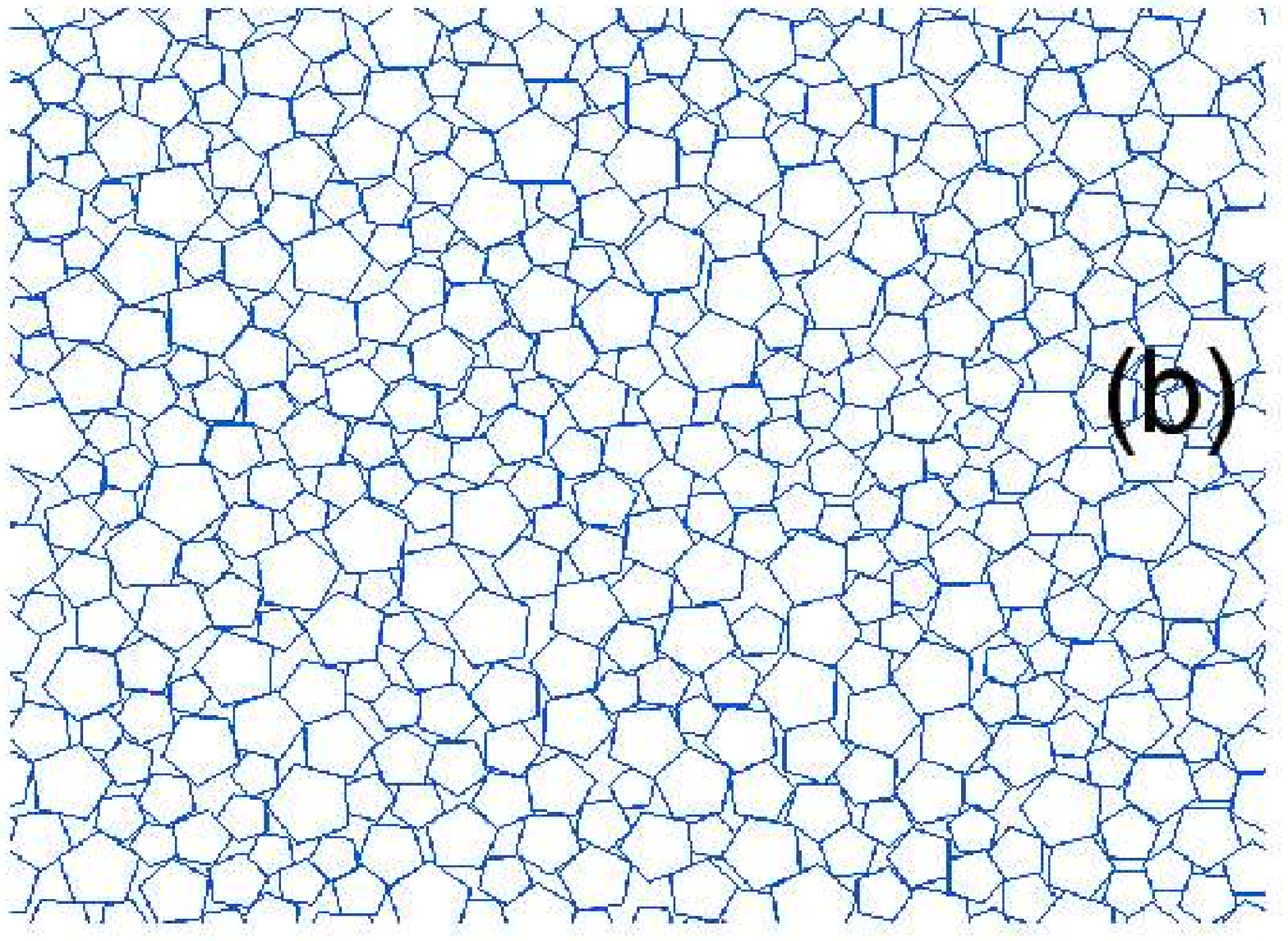}
\caption{Snapshots of a portion of the samples S2 (a) and S1 (b) composed of 
 circular and pentagonal particles, respectively.   \label{fig02}}
\end{figure}

We generated two  numerical samples. The first sample  
S1, is composed of 14400 regular pentagons of three different diameters: 
$50\%$ of diameter $2.5$ cm, $34\%$ of diameter $3.75$ cm and $16\%$ of diameter $5$ cm.
The second sample S2, is composed of 10000 discs 
with the same polydispersity. 
Both samples were prepared according to the same protocol. 
The gravity was set to zero in order to 
avoid force gradients in the samples.
The coefficient of friction was set to 0.4 between grains and to 0 with the walls. 
At equilibrium, both numerical samples were in  isotropic stress state.  
The solid fraction was $\phi_0=0.80$ for S1 and  $\phi_0=0.82$ for S2. 
The aspect ratio was $h/l\approx2$, where $h$ and $l$ are the height and 
width of the sample, respectively. 
Figure \ref{fig02}  displays snapshots of the two packings  
at the end of isotropic compaction. 

The isotropic samples were subjected to vertical compression by downward 
displacement of the top wall at a constant velocity of $1$ cm/s for 
a constant confining stress $\sigma_0$ acting on the lateral walls.
Since we are interested in quasistatic behavior, the shear rate should be 
such that the kinetic energy supplied by shearing is negligible compared to 
the static pressure. 
This can be formulated in terms of an "inertia parameter" $I$  defined by \cite{GDR-MiDi2004}
\begin{equation}
I=\dot \varepsilon \sqrt{\frac{m}{p}},
\label{eq3}
\end{equation}
where $\dot \varepsilon=\dot y /y$ is the strain rate, 
$m$ is the total mass, and $p$ is the average pressure. The quasistatic limit is
characterized by the condition $I\ll1$. In our biaxial simulations, $I$ was 
below $10^{-3}$.  

\section{Shear stress}
\label{sec:3}

In this section, we compare the stress-strain behavior 
between the packings of polygons (sample S1) and disks (sample S2). 
For the calculation of the stress tensor, we consider the "tensorial 
moment" ${\bm M}^i$ of each particle i defined by \cite{Moreau1997,Staron2005}:
\begin{equation}
M^i_{\alpha \beta} = \sum_{c \in i} f_{\alpha}^c r_{\beta}^c,
\label{eq:M}
\end{equation}
where  $f_{\alpha}^c$ is the $\alpha$ component of the force exerted on 
particle i at the contact c, $r_{\beta}^c$ is the $\beta$ component 
of the position vector of the same contact c, and the summation 
is runs over all contacts c of neighboring particles with the particle i (noted briefly by $c \in i$).
It can be shown that the tensorial moment of a collection of rigid particles is the sum of the 
tensorial moments of individual particles. 
The stress tensor ${\bm \sigma}$ for a packing of volume $V$  
is simply given by \cite{Moreau1997,Staron2005}: 
\begin{equation}
{\bm \sigma } = \frac{1}{V} \sum_{i \in V} {\bm M}^i =   \frac{1}{V}  \sum_{c \in V} f_{\alpha}^c \ell_{\beta}^c, 
\label{eq:M}
\end{equation}
where ${\bm \ell}^c$ is the intercenter vector joining the centers of the two touching particles at the 
contact $c$. 
We extract the mean stress $p=(\sigma_1+\sigma_2)/2$, and the stress deviator $q=(\sigma_1-\sigma_2)/2$,
where $\sigma_1$ and $\sigma_2$ are the principal stresses. 
The major principal direction during vertical compression is vertical, 
we  then define the volumetric strain by : 
\begin{equation}
\varepsilon_p =\int_{V_0}^V \frac{dV'}{V'} = \ln \left( 1+ \frac{\Delta V}{V_0} \right),
\end{equation}
where $V_0$ is the initial volume and $\Delta V = V - V_0$ is the cumulative volume change.

\begin{figure}
\centering
\includegraphics[width=8cm]{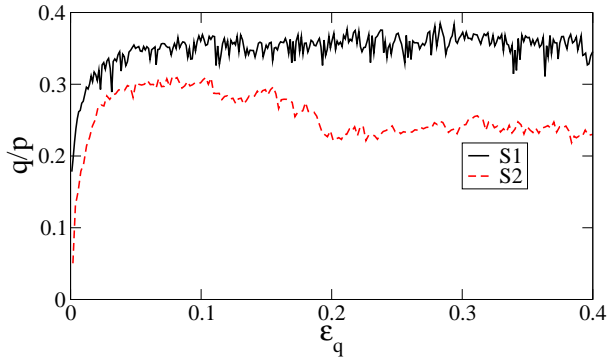}
\caption{Normalized shear stress $q/p$ as a function of cumulative shear strain $\varepsilon_q$ 
for the samples S1 and S2. \label{fig03}}
\end{figure}

Figure \ref{fig03} shows the normalized shear stress $q/p$ for the samples S1 and S2 
as a function of shear  strain  $\varepsilon_q \equiv \varepsilon_1 - \varepsilon_2$. 
For S2, we observe a hardening behavior 
followed by (slight) softening and a stress plateau corresponding to the 
residual state of soil mechanics \cite{Mitchell2005}. For S1, we observe no marked stress peak.  
The residual stress is higher for polygons ($\simeq 0.35$) than for disks ($ \simeq 0.28$). 
The higher level of $q/p$  for the polygon packing reflects the 
organization of the microstructure and the features of force 
transmission that we analyze in more detail below.  

\section{Fabric anisotropy}
\label{sec:4}

The shear strength of dry granular materials is generally attributed to  
the buildup of an anisotropic structure during shear due to 
friction between the particles and as a result of steric effects depending on  
 particle shape \cite{Oda1980,Cambou1993,Radjai2004a}. 

The probability density function $P_\theta (\theta)$, where $\theta$ is the orientation of 
the contact normal $\bm n$,   
provides the basic orientational statistical information  
about the granular the fabric. It is $\pi$-periodic 
in the absence of an intrinsic polarity for  $\bm n$. 
Most lowest-order information is generally given by the 
second moment of $P_\theta$, 
called {\em fabric tensor}  \cite{Satake1982}:  
\begin{equation}
F_{\alpha \beta} = 
\frac{1}{\pi} \int_0^\pi  n_\alpha (\theta) n_\beta (\theta)  P_\theta(\theta) d\theta 
\equiv \frac{1}{N_c} \sum_{c\in V} n_\alpha^c n_\beta^c, 
\label{eq:F}
\end{equation}
where $\alpha$ and  $\beta$ design the components in a reference frame and 
$N_c$ is the 
total number of contacts in  the control volume $V$.  
By definition, $tr ({\bm F}) = 1$. The 
anisotropy of the contact network is given the difference between the principal values 
$F_1$ and $F_2$. We define  the fabric anisotropy $a$ by 
\begin{equation}  
a = 2 (F_1 - F_2).
\end{equation}

\begin{figure} 
  \centering
  \begin{minipage}{5.5cm}
   \hspace*{-0.1cm}
   \includegraphics[width= 5.5cm]{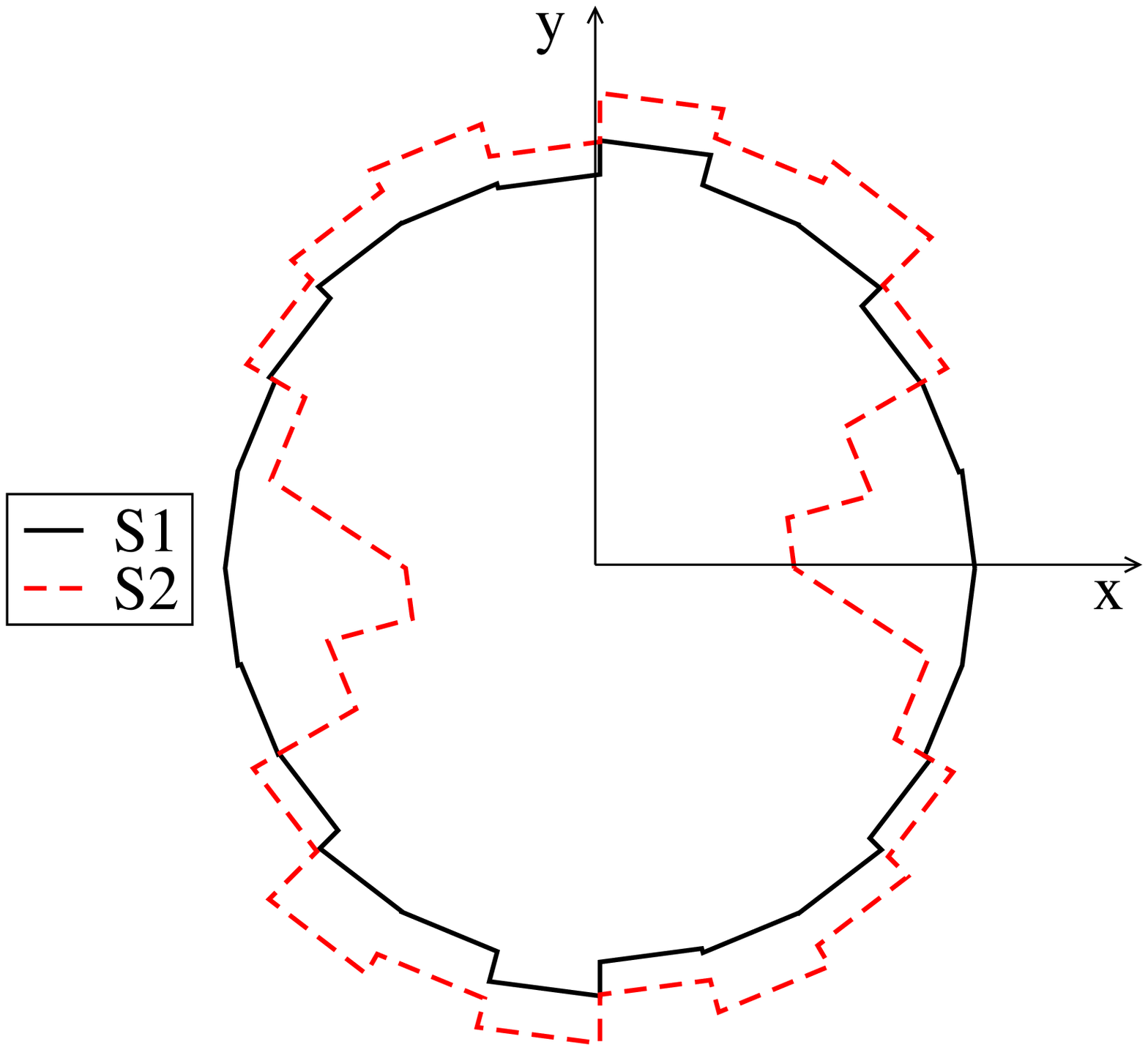}
   \caption{Polar representation of the probability density function 
   $P_\theta$ of the contact normal directions $\theta$ 
   for the samples S1 and S2 in the residual state.     \label{fig08}}
 \end{minipage}
  \hspace*{0.2cm}
  \centering
  \begin{minipage}{5.5cm}
    \hspace*{-0.1cm}
    \includegraphics[width= 5.5cm]{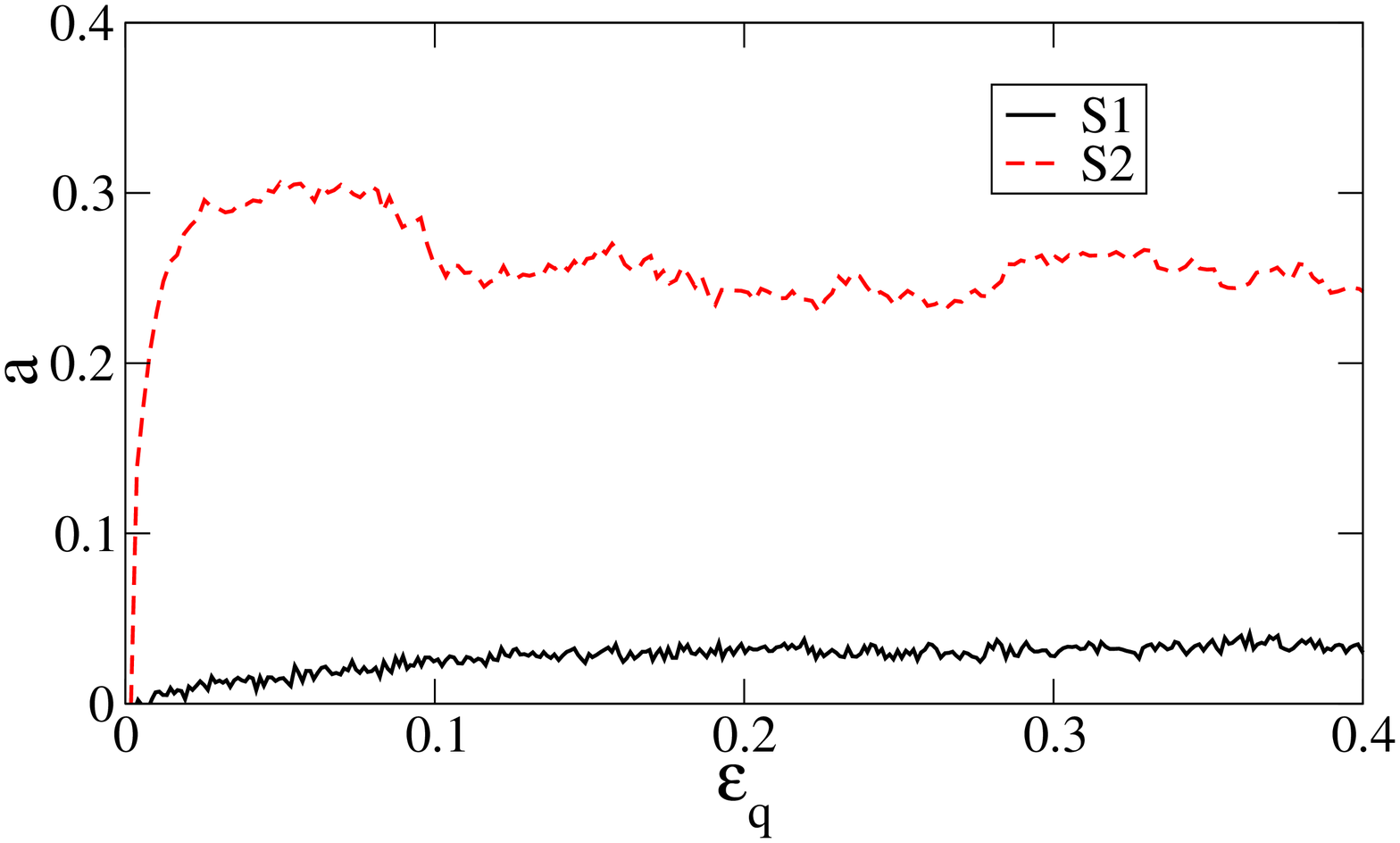}
    \caption{Evolution of the anisotropy $a$ with  cumulative shear strain $\varepsilon_q$
     for the samples S1 and S2.    \label{fig09}}
  \end{minipage}
\end{figure}

Figure \ref{fig08} displays a polar representation of 
$P_\theta(\theta)$ for the samples S1 and S2 at $\varepsilon_q = 0.3$. We observe 
a nearly isotropic distribution for the pentagon packing in spite of shearing whereas the 
disk packing is markedly anisotropic.  This is a surprising observation in view of 
the higher shear strength of the pentagon packing.

The evolution of $a$ is shown in Fig. \ref{fig09} as a function of $\varepsilon_q$ for 
S1 and S2. The anisotropy stays quite weak  in the pentagon packing whereas the disk packing 
is marked by a much larger anisotropy, increasing to $\simeq 0.3$ 
and then relaxing to a slightly lower value in the residual state. 
The low anisotropy of the pentagon packing results from a particular 
organization of the force network in correlation with the orientations of edge-to-edge and vertex-to-edge contacts in 
the packing \cite{azema2007}.  
This leads us to consider the contributions of force and texture anisotropies 
to average shear stresses.

\section{Force anisotropy}
\label{sec:5}
   
The angular distribution of contact forces in a granular packing can be represented by the 
average force ${\langle \bm f \rangle}({\bm n})$ as a function of the contact normal direction $\bm n$. 
We  distinguish the average normal force $\langle f_n \rangle (\theta)$ from the 
average tangential force $\langle f_t \rangle (\theta)$. As $P(\theta)$, 
these two  functions can be represented by their Fourier expansions truncated 
beyond the second term \cite{Kruyt1996,Bathurst1988,Rothenburg1989,Radjai1998}: 
\begin{equation}
\left\{
\begin{array}{lcl}
 \langle f_n \rangle (\theta) &=& \langle f \rangle  \{ 1 + a_n \cos 2(\theta - \theta_n) \}  \\ 
\langle f_t \rangle (\theta) &=&  \langle f \rangle  a_t \sin 2(\theta - \theta_t)
\end{array}
\right.
\label{eqn:fnft}
\end{equation}            
where $\langle f \rangle$ is the average force, $a_n$ and $a_t$ represent the anisotropies of the 
normal and tangential forces, respectively, and  $\theta_n$ and $\theta_t$ are their 
privileged directions.  
It is  convenient to estimate the anisotropies from  the following "force tensors'': 
\begin{equation}
\left\{
\begin{array}{lcl}
H^{(n)}_{\alpha \beta} &=& 
 \int\limits_{0}^\pi  
\langle f_n \rangle(\theta)  n_\alpha  n_\beta d\theta ,  \\ 
H^{(t)}_{\alpha \beta} &=& 
 \int\limits_{0}^\pi  
\langle f_t \rangle(\theta)  n_\alpha  t_\beta d\theta.  \\ 
\end{array}
\right.
\label{eqn:HnHt}
\end{equation}      
Then,  we have $tr({\bm H}^{(n)})= \langle f \rangle$, and 
\begin{equation}
\left\{
\begin{array}{lcl}
a_n &=& 2 \frac{H^{(n)}_1  - H^{(n)}_2}{H^{(n)}_1  + H^{(n)}_2}, \\
a_t &=& 2 \frac{H^{(t)}_1  - H^{(t)}_2}  {H^{(n)}_1  + H^{(n)}_2}, \\
\end{array}
\right.
\label{eqn:anat}
\end{equation}  
where the subscripts $1$ and $2$ refer to the principal values of the tensors.  

\begin{figure}
\centering
\includegraphics[width=5.5cm]{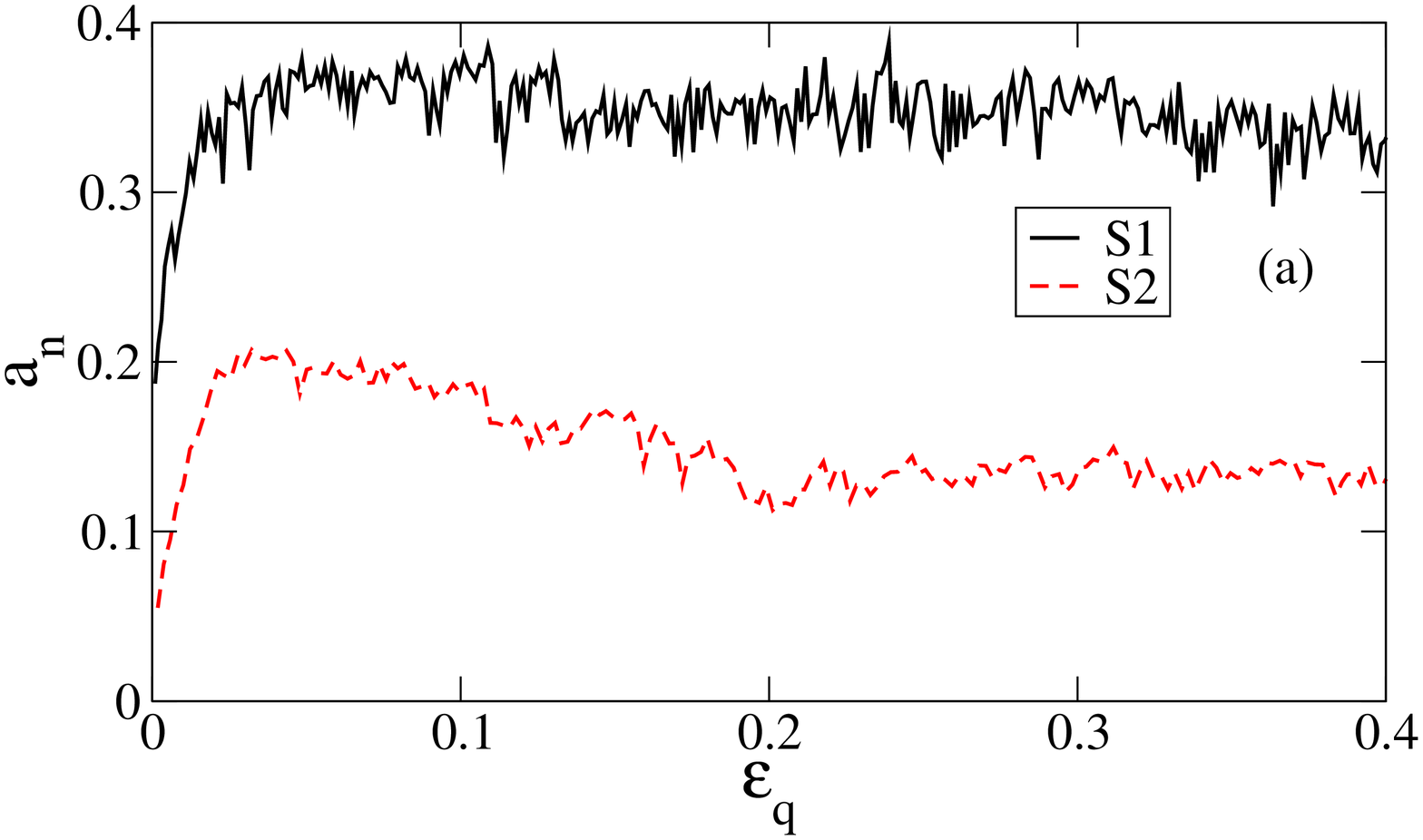}
\includegraphics[width=5.5cm]{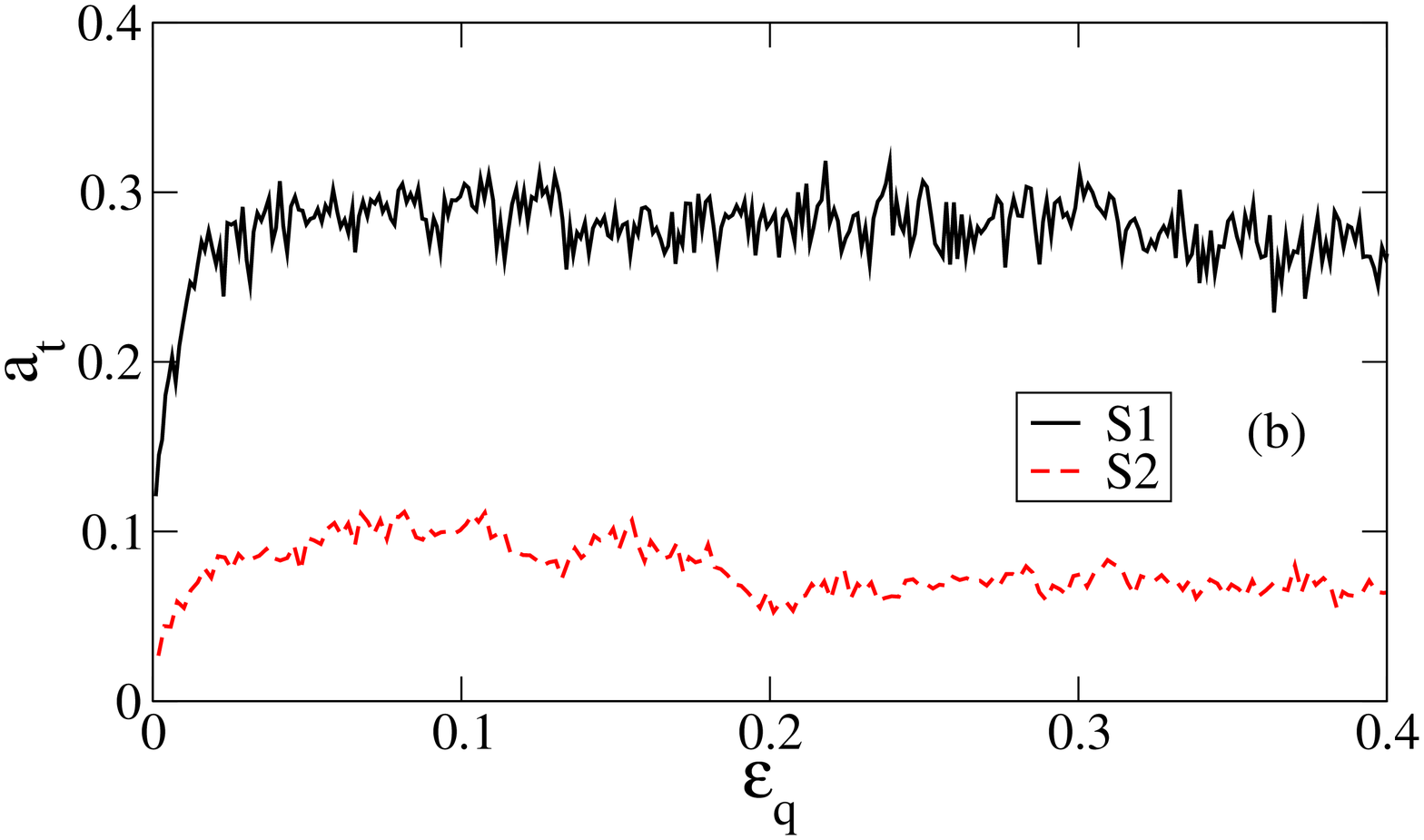}
\caption{Evolution of force anisotropies  $a_n$ (a) and $a_t$ (b) as a function 
of cumulative shear strain  $\varepsilon_q$ in samples S1 and S2.  \label{fig11}}
\end{figure}
\begin{figure}
\centering
\includegraphics[width=5cm]{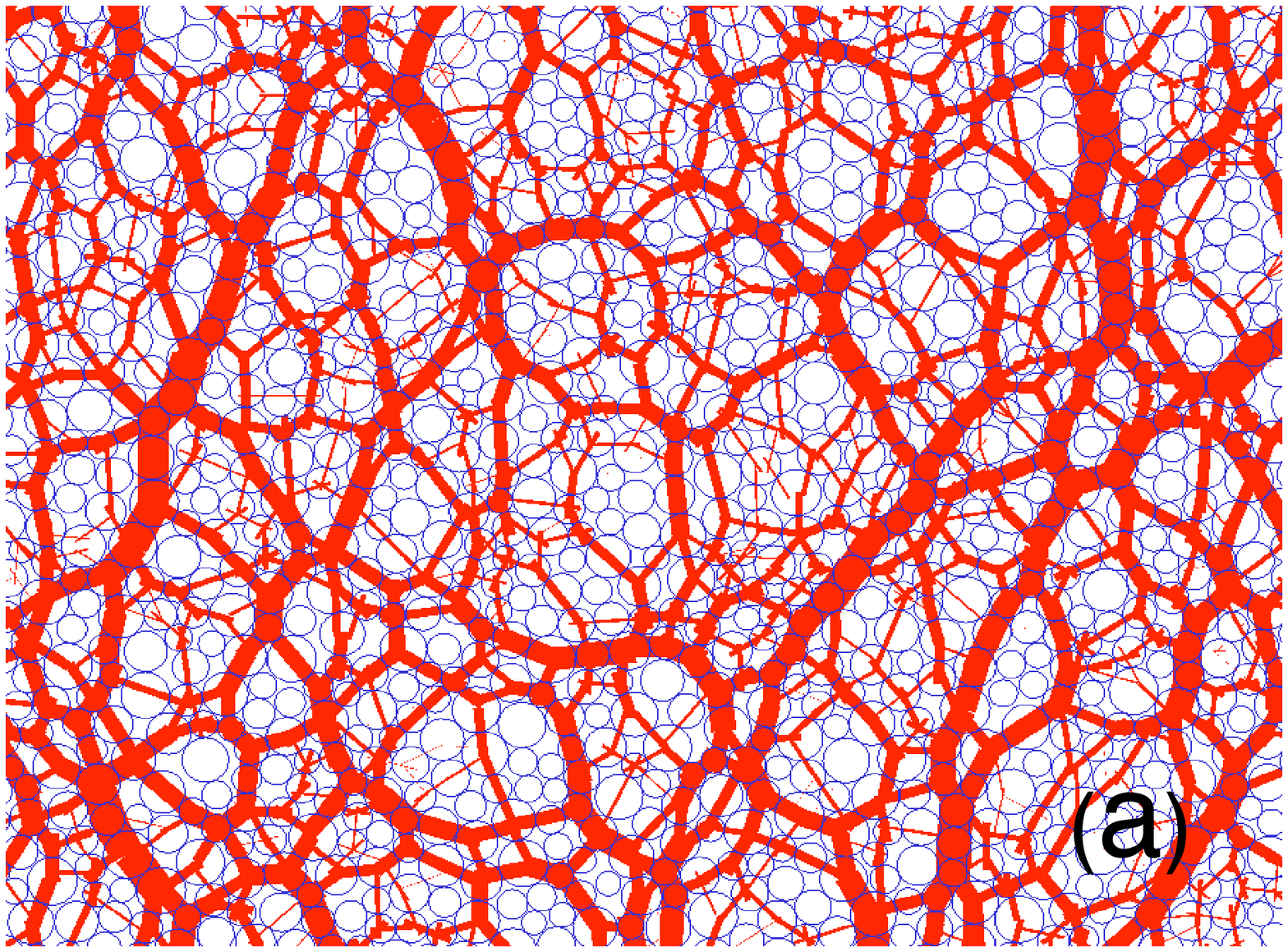}
\includegraphics[width=5cm]{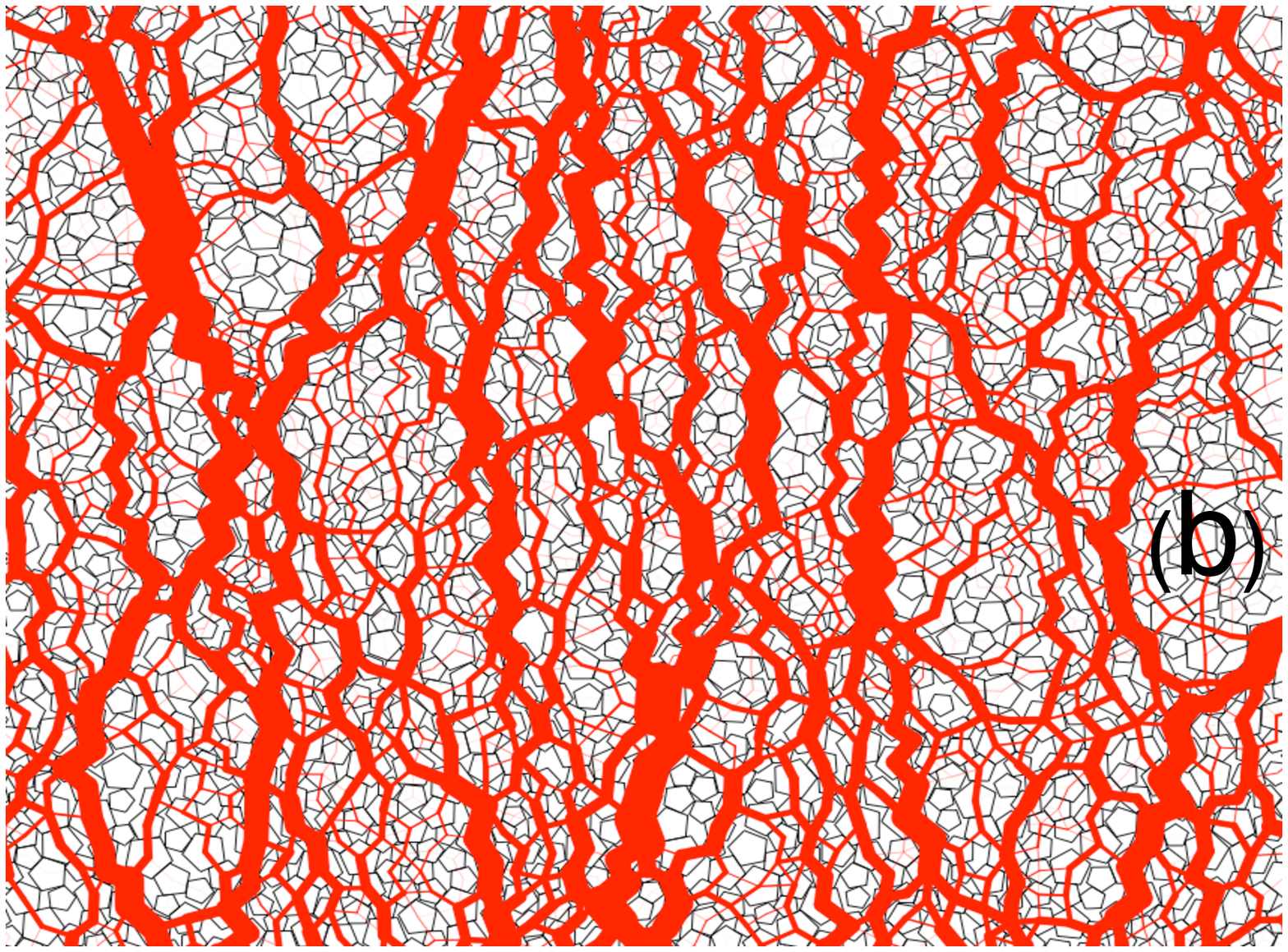}
\caption{Snapshots of normal forces in samples S2 (a) and S1 (b). Line thickness is proportional 
to the normal force. \label{fig13}}
\end{figure}

Figure \ref{fig11} 
shows the evolution of $a_n$ and $a_t$ with $\varepsilon_q$ in samples S1 and S2. We see that, 
in contrast to fabric anisotropies (Fig. \ref{fig09}), 
the force anisotropies in the pentagon packing remain always above those in 
the disk packing. This means that the aptitude of the pentagon packing to develop 
large force anisotropy and strong force chains 
is not solely dependent on the global fabric anisotropy of the system. Indeed, due to 
the geometry of the pentagons, i.e. the absence of parallel sides, the strong force chains 
are mostly of zig-zag shape, as observed in Fig. \ref{fig13}, and the 
stability of such structures requires strong activation of tangential forces. This 
explains, in turn, the large value of $a_t$ for pentagons, very close to $a_n$, whereas in the disk packing 
$a_t$ is nearly half of $a_n$.

\begin{figure}
\centering
\includegraphics[width=8cm]{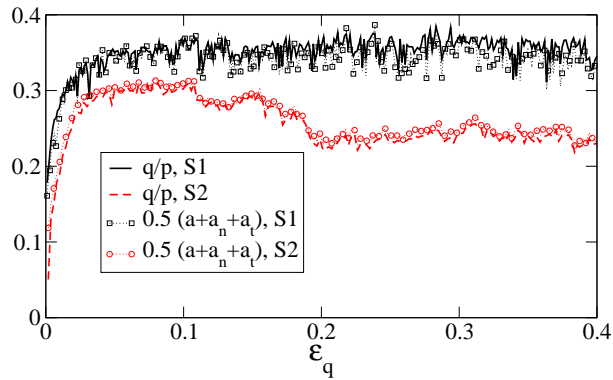}
\caption{Evolution of the normalized shear stress  $q/p$ for the samples S1 and S2 
with $\varepsilon_q$ 
together with the corresponding predictions from its expression as a function of 
fabric and force anisotropies Eq. (\ref{eq:qa}).\label{fig12}}
\end{figure}

The  anisotropies $a$, $a_n$ and $a_t$ are interesting descriptors of granular 
microstructure and force transmission as they underlie the 
shear stress. Indeed,     
it can be shown that the general expression of the stress tensor Eq. (\ref{eq:M}) under some 
approximations leads to the 
following simple relation \cite{Rothenburg1989,Radjai2004a}:
\begin{equation}     
\frac{q}{p} \simeq \frac{1}{2} (a+a_n+a_t),
\label{eq:qa}
\end{equation}
where the cross products $a_n a$ and $a_t a$ between the anisotropies have been neglected compared to the anisotropies, and it has been assumed that the stress tensor is coaxial with the fabric tensor Eq. (\ref{eq:F}) and the force tensors Eq. (\ref{eqn:HnHt}). 
Fig. \ref{fig12} shows that Eq. (\ref{eq:qa}) holds quite well both for pentagons and 
disks. This equation provides an amazingly good estimate of the shear stress from 
the anisotropies under monotonic shearing.

A remarkable consequence of Eq. (\ref{eq:qa}) is to reveal 
the distinct origins of shear stress in pentagon and disk packings. 
The fabric anisotropy provides a major contribution to  shear stress in the disk packing 
(Fig. \ref{fig09}) whereas 
the force anisotropies are more important for shear stress in 
the pentagon packing (Fig. \ref{fig11}). In this way, in spite of the weak fabric anisotropy $a$, 
the larger force anisotropies $a_n$ and $a_t$ allow 
the pentagon packing  to reach higher levels of $q/p$ compared to the disk packing.            

\section{Conclusion}

The objective of this paper was to isolate the effect of particle shape 
on shear strength in granular media by comparing
two similar packings with different particle shapes: pentagons vs. disks. 
We observed enhanced shear strength and force inhomogeneity in 
the pentagon packing. But, unexpectedly, the pentagon packing was found to develop 
a lower structural (fabric) anisotropy compared to the disk packing under shear. 
This low fabric anisotropy, however,  does not  prevent the pentagon packing from  
building up a strong force anisotropy that underlies its enhanced shear strength 
compared to the disk packing. 

This finding is interesting as it shows  that the force anisotropy in a 
granular material depends on the particle shapes. 
This mechanism may be the 
predominant source of strength for "facetted" particles that can give rise edge-to-edge (in 2D) 
contacts allowing for strong force localization along such contacts 
in the packing.  
Since the fabric anisotropy is low in a pentagon packing, the role of 
force anisotropy and thus the local equilibrium structures are important 
with respect to its overall strength properties.  
With pentagon packings, we were able to demonstrate the nontrivial 
phenomenology resulting from the specific shape of particles  as compared to 
a disk packing. We found a similar behaviour for other   
 regular polygons (hexagons and higher number of sides) as 
well as polyhedral particles in three dimensions.

\printindex

\begin{thebibliography}{10}

\bibitem{Liu1995a}
{C.-h.} Liu, S.~R. Nagel, D.~A. Schecter, S.~N. Coppersmith, S.~Majumdar,
  O.~Narayan, and T.~A. Witten.
\newblock Force fluctuations in bead packs.
\newblock {\em Science}, 269:513, 1995.

\bibitem{Radjai1996}
F.~Radjai, M.~Jean, J.J. Moreau, and S.~Roux.
\newblock Force distributions in dense two dimensional granular systems.
\newblock {\em Phys. Rev. Letter}, 77:274--277, 1996.

\bibitem{Coppersmith1996}
S.~N. Coppersmith, {C.-h.} Liu, S.~Majumdar, O.~Narayan, and T.~A. Witten.
\newblock Model for force fluctuations in bead packs.
\newblock {\em Phys. Rev. E}, 53(5):4673--4685, 1996.

\bibitem{Mueth1998a}
D.~M. Mueth, H.~M. Jaeger, and S.~R. Nagel.
\newblock Force distribution in a granular medium.
\newblock {\em Phys. Rev. E.}, 57(3):3164--3169, 1998.

\bibitem{Lovol1999}
G.~Lovol, K.~Maloy, and E.~Flekkoy.
\newblock Force measurments on static granular materials.
\newblock {\em Phys. Rev. E}, 60:5872--5878, 1999.

\bibitem{Bardenhagen2000}
S.~G. Bardenhagen, J.~U. Brackbill, and D.~Sulsky.
\newblock Numerical study of stress distribution in sheared granular material
  in two dimensions.
\newblock {\em Phys. Rev. E}, 62:3882--3890, 2000.

\bibitem{Antony2001}
S.~J. Antony.
\newblock Evolution of force distribution in three-dimensional granular media.
\newblock {\em Phys Rev E}, 63:011302, 2001.

\bibitem{Majmudar2005}
T.~S. Majmudar and R.~P. Behringer.
\newblock Contact force measurements and stresse-induced anisotropy in granular
  materials.
\newblock {\em Nature}, 435:1079--1082, 2005.

\bibitem{Silbert2002}
L.~E. Silbert, G.~S. Grest, and J.~W. Landry.
\newblock Statistics of the contact network in frictional and frictionless
  granular packings.
\newblock {\em Phys. Rev. E}, 66:1--9, 2002.

\bibitem{Radjai1998}
F.~Radjai, D.~E. Wolf, M.~Jean, and J.J. Moreau.
\newblock Bimodal character of stress transmission in granular packings.
\newblock {\em Phys. Rev. Letter}, 80:61--64, 1998.

\bibitem{Kruyt1996}
N.~P. Kruyt and L.~Rothenburg.
\newblock Micromechanical definition of strain tensor for granular materials.
\newblock {\em ASME Journal of Applied Mechanics}, 118:706--711, 1996.

\bibitem{Bathurst1988}
R.~J. Bathurst and L.~Rothenburg.
\newblock Micromechanical aspects of isotropic granular assemblies with linear
  contact interactions.
\newblock {\em J. Appl. Mech.}, 55:17, 1988.

\bibitem{Rothenburg1989}
L.~Rothenburg and R.~J. Bathurst.
\newblock Analytical study of induced anisotropy in idealized granular
  materials.
\newblock {\em Geotechnique}, 39:601--614, 1989.

\bibitem{Staron2005}
L.~Staron and F.~Radjai.
\newblock Friction versus texture at the approach of a granular avalanche.
\newblock {\em Phys. Rev. E}, 72:1--5, 2005.

\bibitem{Jean1999}
M.~Jean.
\newblock The non smooth contact dynamics method.
\newblock {\em Computer Methods in Applied Mechanic and Engineering},
  177:235--257, 1999.

\bibitem{Jean1992}
M.~Jean and J.~J. Moreau.
\newblock Unilaterality and dry friction in the dynamics of rigid body
  collections.
\newblock In {\em Proceedings of Contact Mechanics International Symposium},
  pages 31--48, Lausanne, Switzerland, 1992. Presses Polytechniques et
  Universitaires Romandes.

\bibitem{Moreau1994}
J.J. Moreau.
\newblock Some numerical methods in multibody dynamics : application to
  granular.
\newblock {\em Eur. J. Mech. A/Solids}, 13:93--114, 1994.

\bibitem{DUBOIS2003}
F.~Dubois and M.~Jean.
\newblock Lmgc90 une plateforme de d\'eveloppement d\'edi\'ee \`a la
  mod\'elisation des probl\`emes d'int\'eraction.
\newblock In {\em Actes du sixi\`eme colloque national en calcul des structures
  - CSMA-AFM-LMS -}, volume~1, pages 111--118, 2003.

\bibitem{GDR-MiDi2004}
GDR-MiDi.
\newblock On dense granular flows.
\newblock {\em Eur. Phys. J. E}, 14:341--365, 2004.

\bibitem{Moreau1997}
J.~J. Moreau.
\newblock Numerical investigation of shear zones in granular materials.
\newblock In D.~E. Wolf and P.~Grassberger, editors, {\em Friction, Arching,
  Contact Dynamics}, pages 233--247, Singapore, 1997. World Scientific.

\bibitem{Mitchell2005}
J.K. Mitchell and K.~Soga.
\newblock {\em Fundamentals of Soil Behavior}.
\newblock Wiley, NY, 2005.

\bibitem{Oda1980}
M.~Oda, J.~Koshini, and S.~Nemat-Nasser.
\newblock Some experimentally based fundamental results on the mechanical
  behavior of granular materials.
\newblock {\em Geotechnique}, 30:479--495, 1980.

\bibitem{Cambou1993}
B.~Cambou.
\newblock From global to local variables in granular materials.
\newblock In C.~Thornton, editor, {\em Powders and Grains 93}, pages 73--86,
  Amsterdam, 1993. A. A. Balkema.

\bibitem{Radjai2004a}
F.~Radjai, H.~Troadec, and S.~Roux.
\newblock Key features of granular plasticity.
\newblock In S.J. Antony, W.~Hoyle, and Y.~Ding, editors, {\em Granular
  Materials: Fundamentals and Applications}, pages 157--184, Cambridge, 2004.
  RS.C.

\bibitem{Satake1982}
M.~Satake.
\newblock Fabric tensor in granular materials.
\newblock In P.~A. Vermeer and H.~J. Luger, editors, {\em Proceedings of the
  IUTAM symposium on deformation and failure of granular materials, Delft},
  pages 63--68, Amsterdam, 1982. A. A. Balkema.

\bibitem{azema2007}
E.~Az\'ema, F.~Radjai, R.~Peyroux, and G.~Saussine.
\newblock Force transmission in a packing of pentagonal particles.
\newblock {\em Phys. Rev. E}, 76:011301, 2007.

\end{thebibliography}
\end{document}